# The Effects of Socioeconomic Status and Depression on The Neural Correlates of Error Monitoring. An Event-Related Potential Study

Hiran Shanake Perera*[1],

Rozainee Khairudin[1],

Khazriyati Salehuddin[2]

[1]School of Liberal Arts and Sciences, Department of Psychology, Taylor's University, Subang Jaya, Malaysia.

[2]Faculty of Social Sciences and Humanities, Universiti Kebangsaan Malaysia.

*Corresponding author: hiranpererawa@outlook.com

## Abstract

Existing evidence suggests that neural responses to errors were exaggerated in individuals at risk of depression and anxiety. This phenomenon has led to the possibility that the error-related negativity (ERN), a well-known neural correlate of error monitoring could be used as a diagnostic tool for several psychological disorders. However, conflicting evidence between psychopathology and the ERN suggests that this phenomenon is modulated by variables are yet to be identified. Socioeconomic status (SES) could potentially play a role in the relationship between the ERN and psychopathological disorders, given that SES is known to be associated with depression and anxiety. In the current study, we first tested whether SES was related to ERN amplitude. Second, we examined whether the relationship between the ERN and depression was explained by differences in SES. We measured error-related negativity (ERN) from a sample of adult participants from low to high socioeconomic backgrounds while controlling their depression scores. Results show that SES correlated with variations in ERN amplitude. Specifically, we found that low-SES individuals had a larger ERN than wealthier individuals. In addition, the relationship between depression and the ERN was fully accounted for by variations in SES. Overall, our results indicate that SES predicts neural responses to errors. Findings also indicate that the link between depression and ERN may be the result of SES variations. Future research examining the links between psychopathology and error monitoring should control SES differences, and caution is needed if they are to be used as a diagnostic tool in low-income communities.

**INTRODUCTION**

The ability to flexibly detect our own errors is a core aspect of human behaviour, as it is a key determinant of how people learn novel information and adapt to complex situations (Botvinick, Matthew et al., 2001; Hajcak & Simons, 2002; Holroyd & Coles, 2002; Nieuwenhuis & Yeung, 2003). An important development in the field of error monitoring has been the discovery that the error-related negativity (ERN) correlates with individual differences linked to psychopathological disorders (Olvet & Hajcak, 2008). The ERN is a scalp event-related brain potential (ERP) observed on fronto-central electrodes within 100 milliseconds following the commission of an error. This signal is widely seen as a reliable neural index of error monitoring and it is thought to reflect activity in the anterior cingulate cortex (ACC) (Dehaene et al., 1994; Gehring et al., 1993, 2018; Kim et al., 2007; Taylor et al., 2018; Van Veen & Carter, 2002). Many studies linking the ERN and psychopathology have reported that depression levels predict larger ERN amplitudes (Aarts et al., 2013; Holmes & Pizzagalli, 2010; Klawohn et al., 2020; Ruchensky et al., 2020; Tang et al., 2013). In addition, several studies have also reported a pattern of enhanced ERN amplitude in individuals reporting symptoms of anxiety and Obsessive-Compulsive disorders (Carrasco et al., 2013; Hajcak & Simons, 2002) and anxiety (Gorka et al., 2017; ; Wu et al., 2019). (Gorka et al., 2017; Schrijvers et al., 2010; Wu et al., 2019)

Because of these findings, the ERN is now widely regarded as a transdiagnostic index of *internalizing* psychopathological disorders (Gorka et al., 2017; Olvet & Hajcak, 2008), a category that includes a wide range of mood disorders such as depressive disorders, anxiety, and OCD. From this perspective, an enlarged ERN has been interpreted as an exaggerated sensitivity to errors in individuals at risk of these disorders (Hajcak & Simons, 2002), which is consistent with several theoretical models (Clayson et al., 2020)..

However, a number of contradictory findings have also been reported. For instance, many studies have failed to find an association between depression and an enlarged ERN (Olvet et al., 2010; Ruchsow et al., 2006; Schrijvers et al., 2010). A recent meta-analysis has found weak evidential support for the relationship between depression and the ERN (Clayson et al., 2020), and another meta-analysis found a relationship between depression and the ERN, although it also found evidence of publication bias (Moran et al., 2017). In addition, a recent study reported an absence of relationship between anxiety dimensions and the ERN, suggesting that this relationship is less reliable than previously thought, at least in non-clinical samples (Härpfer et al., 2020). These contradictory results suggest that the link between the ERN and psychopathology is not yet fully understood. A potential approach to address this problem is to identify variables that may be confounding or masking this relationship.

This article attempts to tackle this issue by focusing on whether a key individual difference – socioeconomic status (SES) may be playing a role in how psychopathology is related to the ERN. SES is a construct that often refers to differences in income and wealth, but also to subjective positions in society and levels of education (Braveman et al., 2005; Perera-W.A. et al., 2021; Siwar et al., 2016). SES has a strong potential to play a role in the relationship between psychopathology and the ERN for two reasons: First, SES is strongly correlated with individual differences linked to psychopathology. For instance, extensive evidence shows that SES is inversely related to depression scores (Lorant et al., 2003; Lupien et al., 2001; Ridley & Patel, 2020) and a high prevalence of depression and associated psychopathological disorders is typically found in groups living in poverty (Heflin & Iceland, 2009). Evidence also indicates that several anxiety disorders are more prevalent in low-SES communities (Santiago et al., 2011). Second, substantial evidence shows that low-SES individuals tend to overreact to aversive stimuli (Hao & Farah, 2020),  and it has been suggested that this may be caused by the fact that low-SES individuals have a history of living in an overall more punitive environment (Gonzalez et al., 2016). Therefore, it is possible that SES may be linked to an overreaction to errors per se – which are fundamental instances of aversive stimuli (Hajcak & Foti, 2008).

Although studies explicitly investigating how SES relates to the ERN are rare, two studies focusing on very young children  have examined this question. Conejero et al. (2016) observed that toddlers (16-18 months old) from a low-SES background exhibited a reduced ERN amplitude to incorrect responses compared to high-SES toddlers. Brooker (2018) found that ERN amplitude increased over time (from 3 to 4 year-old) for high-SES children but not for low-SES children (Brooker, 2018). These studies are valuable to depict the relationship between SES and error monitoring at early stages of development. However, it is still largely unknown whether SES predicts the ERN in adults whose cognitive and emotional systems have been shaped by multiple life events and socioeconomic factors.

Therefore, this study had two main goals. First, we examined whether SES correlated with ERN amplitude in adults. Given that previous research shows that low-SES individuals are more prone to psychopathological disorders and more likely to overreact to aversive stimuli, we hypothesized that SES would be inversely correlated with ERN amplitude. Second, we tested whether SES could account for the relationship between the ERN and the Center for Epidemiologic Studies Depression Scale (CES-D, Radloff, 1977), a well-known measure of depression which also correlates with other internalizing disorders (Abramowitz et al., 2009; Kavish et al., 2020; Orme et al., 1986). . Here we could test two competing hypotheses: First, if ERN is a reflection of individual differences in ID, then the relationship between CES-D scores and the ERN should be significant after controlling for SES.

However, if the relationship between ID and the ERN is a by-product of a relationship between the ERN and factors associated to SES other than psychopathology, then the link between CES-D and the ERN should be fully accounted for by SES variability. To tackle these questions, we used a portable EEG system to measure the ERN from adults performing a classical Go/No-Go task (Amodio et al., 2007; Braver et al., 2001) in an urban center from a developing country with a large socioeconomic variability – Kuala Lumpur in Malaysia. This sample was selected to include both low-SES and middle to high SES individuals from whom we were able to record the CES-D scores, and family income information as a proxy for SES.

## METHODS

### Participants

Eighty-four right-handed adults with no history of neurological conditions were recruited for this study. Data from 13 participants were excluded because they did not have enough artifact-free trials (see the *Electrophysiological data recording and pre-processing* section). In addition, we detected 2 multivariate outliers in our data using the Mahalanobis distance technique (Leys et al., 2018). The final sample included 69 participants (46 females) with a mean age of 41.06 years (SD = 12.52 years). To obtain a socioeconomically diverse sample, we recruited approximately half of our sample (n = 35) from low-income households, whereas the remainder of the sample came from middle-to high income households. It is important to point out that we recruited our participants from a large urban setting (Kuala Lumpur) in Malaysia, a middle-income developing country with relatively high levels of income inequality (Department of Statistics, 2020.; Human Development Report, 2019; Khazanah Research Institute, 2018). Specifically, low-SES participants were recruited from households identified by the Malaysian Ministry of Women, Family, and Community Development as being at high risk of poverty.

Accordingly, we recorded a median monthly household income of MYR1500.50 (USD 359.62) from the 35 low-SES participants, which is below the Malaysian 2020 poverty line (MYR2208, USD 534) (Program Pembasmian Kemiskinan Bandar (PPKB), 2020). Middle to high income participants (n = 34) were recruited via social media, word of mouth and snowballing, and they reported a median monthly household income of MYR5500.50, (USD 1318.28). Both low and high SES were matched for gender (70% females), ethnicity (90% Malay), and age (low-SES had a mean age of 41.46, and high SES had a mean age of 38.85, p = 0.36. Upon completing the study, participants received a token financial reward (RM60 approx. 14.50 USD equivalent) to thank them for their participation and they all signed an informed consent. The study was approved by the Ethics committee of the Universiti Kebangsaan Malaysia (The National University of Malaysia), one of the main public research universities in Malaysia (Ref: JEP-2018-339).

**Self-report Measures**

***Socioeconomic Status.***

Consistent with previous research (Brooker, 2018; Conejero et al., 2016; Kishiyama et al., 2009), we used Family income to operationalize socioeconomic status (SES). All participants reported their monthly household income on a 9-point Likert-scale, which provided a fine-grained breakdown of income at the low and high levels (1 = MYR1000 or less, 2 = MYR1001 – MYR2000, 3 = MYR2001 – MYR3000, etc.). Although our sample could be divided into two *a priori* income categories (low and high family income) we treated family income as a continuous variable in most of our statistical analyses to minimize information loss that might mask important statistical effects (MacCallum et al., 2002). We do report a dichotomous treatment of how SES predicts ERN amplitude in Figure 2c to facilitate the visualization and understanding of our data. In the remainder of this article, SES will refer to family income.

***CES-D.***

To measure common depressive symptoms, we used the Center for Epidemiologic Studies Depression Scale (CES-D, Radloff, 1977). This questionnaire consists of twenty items on 4-point Likert scales with total scores ranging from 0 to 60. The CES-D has been widely used in both research and clinical settings; it is thought to have strong psychometric properties (Radloff, 1977; Vilagut et al., 2016) and it also correlates with other internalizing disorders such as anxiety, obsessive-compulsive disorders and psychopathy (Orme et al., 1986; Kavish et al., 2020; Abramowitz et al., 2009). We used an existing validated translation in *Bahasa Malaysia*, the national language of Malaysia (Mazlan & Ahmad, 2014). The Cronbach alpha for our sample was of 0.92. Reliability figures for each CESD subscale were high: Depressed affect (DA) factor, $\alpha = .89$; Somatic symptoms (SS), $\alpha = .80$; Positive affect (PA), $\alpha = .79$. Our high SES group had a median score of 17.5 and our low-SES group had a median score of 25, which is above the recommended cut-off score of 20 (Vilagut et al., 2016). This is consistent with previous research showing a high risk of depression for individuals living in poverty (Lorant et al., 2003; Lupien et al., 2001). However, it has to also be pointed out that existing research suggests that higher CESD cut-off scores should be used for samples drawn from Malaysia in order to avoid false positives (Ghazali et al., 2014).

**Procedure and design**

Participants were tested in a community room of the Ministry of Women, Family, and Community Development (*Kementerian Pembangunan Wanita, Keluarga, dan Masyarakat*, KPWKM), in central Kuala Lumpur. The room was well-lit and isolated from external noises. Upon arrival, participants were first given the informed consent form, a questionnaire assessing their socioeconomic status and sociodemographic information, and CES-D. Questionnaires were administered following a structured interview approach in order to minimize effects of literacy and to make sure that all the questions were fully understood. Next, participants were prepared for the recording of EEG activity.

Consistent with previous research we used a Go/No-Go (GNG) task to elicit the ERN, (e.g. Amodio et al., 2007; Nieuwenhuis et al., 2003). The task was prepared with PsychoPy version 3.5 (Peirce, 2007), and displayed on a 21" screen. Participants responded with an attached external QWERTY keyboard. The experimenter was present throughout the testing; however, no feedback was provided to the participants throughout the duration of the experiment. Each participant sat approximately 75 cm away from the computer screen with the keyboard placed in front of them. Correct Go responses were recorded by pressing the "SPACE" key on an attached keyboard.

The task  was explained to the participants before the beginning of the practice trials. Participants responded (via a keypress) rapidly to a frequently occurring stimulus known as the "Go" response and refrain from responding to the infrequently occurring "No-Go" stimulus that occurs on a smaller portion of trials. There were 20 practice trials followed by 350 trials (80% of Go and 20% of No-Go trials) separated by 3 short breaks. The behavioural task lasted on average 12-minutes. Each trial started with a 500 ms fixation cross on a grey background followed by the target letter "O" or "X" presented at the center of the screen for 200 ms. In each experimental group, half of the participants were instructed to press a "SPACE" key for the Go-response when they saw "O" on the screen, and not to press any key when they saw "X". The remaining participants pressed the "SPACE" key for "X" for Go-response, and no response for "O". Participants' assignment for each version was random. Participants were instructed to press the "SPACE" key as soon as they saw the target go stimulus. After the target, a blank screen appeared for 600 ms followed by a performance feedback for 500 ms. Participants were instructed to respond within 600 ms of target onset. If the response was slow (> 600 ms), a "Too Slow" warning message appeared, or an "Incorrect" message appeared if the response was wrong. On average, participants in the low SES group answered 305.55 trials correctly (SD = 30.82) and 44.45 (SD = 30.82) trials incorrectly. The high SES participants on average answered 309.82 (SD = 26.66) trials correctly, and 40.17 (SD = 26.66) trials incorrectly.

**Electrophysiological data recording and pre-processing**

Continuously recorded EEG was acquired from a Cognionics HD-72 high-impedance 32-channels dry and wireless EEG headset (Cognionics, Inc. San Diego, CA, USA, Chi et al., 2013; Mullen et al., 2015) at a rate of 500 Hz, a recording bandwidth of 0.1-250Hz and an impedance ≤100 kΩ. Electrode locations fitted the 10-20 system. The Cognionics system is equipped with an active ground system and a Faraday-cage like enclosure to minimize interference from electrical noise (Chi et al., 2013; Mullen et al., 2015), which makes it ideal for EEG data collection taking place outside of a controlled lab environment.

Raw EEG data were recorded with a left mastoid reference, and re-referenced offline to a common average reference. EEG data were pre- processed using EEGLAB 14.1.1b (Delorme & Makeig, 2004) and ERPLAB version 6.1.4 (Lopez-Calderon & Luck, 2014). Data were filtered offline (1−30 Hz), segmented into epochs between -500ms before and 500 ms after the onset of each response. Data was baseline corrected using a -400 to -200 baseline epoch (Härpfer et al., 2020; Olvet et al., 2010; Wang et al., 2015).

Independent Component Analysis (ICA) was run on epoched data using the "Infomax" ICA decomposition method implemented in the "runica" function of EEGLAB (Delorme & Makeig, 2004). We next followed standard guidelines (Jung et al., 2000) to detect and remove components accounting for ocular and muscular artifacts. Next, we rejected data epochs with max-min amplitudes exceeding 100μV in 1000-ms segments isolated in steps of 100 ms. In line with recommended practice (Picton et al., 2000), we excluded 13 participants from the initial sample  because they had more than one third of their trials rejected according to these criteria. Five of those participants also had less than 6 artifact-free error trials. On average, 13.63% (SD = 7.15%) of trials were removed from each remaining participant following these methods (with a maximum of 26.5%). The average number of artifact-free correct go trials were 167, and incorrect No-Go trials were 66. The minimum number of artifact-free trials was set at 6, consistent with previous recommendations for ERN research (Olvet & Hajcak, 2009). On average, 3.01 channels were found to be artifactual. They were interpolated either through spherical spline interpolation or nearest-neighbour replacement.

Correct and error trials were averaged separately to calculate response-locked ERPs. Following previous practice (Amodio et al., 2007; Gehring et al., 1995; Weinberg et al., 2016). Both the ERN and CRN were quantified as mean amplitudes extracted from a window between -80 ms before and 100 ms after the response onset at the *Fz* electrode. The CRN amplitude was also calculated at the same time window following a correct response. We then calculated the ΔERN, the difference between the ERN and CRN, by subtracting the activity of correct go trials from Incorrect No-Go trials (Meyer et al., 2018; Peters et al., 2019; Yeung et al., 2004). To test our main hypotheses, a series of correlations and multiple regression analyses were computed in which ΔERN was the main dependent variable, and SES and CESD were the main predictors. We also provide results obtained from individual ERN and CRN waveforms in the supplementary section (See Tables S2 and S3). Split half reliability was determined by Spearman-Brown corrected correlations of odd- and even-numbered trials to examine the psychometric properties of the ERN ($r$ = .76), and CRN ($r$ = .75).

Finally, we also analyzed the error positivity (Pe), a component often associated with the ERN (Falkenstein et al., 2000). We extracted mean amplitudes between 200 and 400 ms after stimulus onset (Klawohn et al., 2020) from an average of 3 posterior electrodes (P3, Pz and P4). This choice was guided by previous research showing that the Pe is often maximal in posterior electrodes (Falkenstein et al., 2000) and from a careful visual inspection of our waveforms. We next computed a difference waveform (ΔPe) between the incorrect and correct Pe waveforms. All other parameters were similar to the quantification of the ERN.

# RESULTS

## Behavioural and self-report data

We computed correlations to investigate the relationships between SES, CES-D scores and task performance (Table 1). SES inversely correlated with CES-D, and increasing SES predicted a lower error rate and slower response times in the Go/No-Go task, which is consistent with previous literature of how SES relates to depression and cognitive function (Heflin & Iceland, 2009; Klawohn et al., 2020; Lupien et al., 2001; Ridley,& Patel, 2020; Tang et al., 2013).

Consistent with previous research (Amodio et al., 2007; Kim et al., 2007; D. Schrijvers et al., 2008),, 30% errors were committed on No-Go trials (commission errors) and 0.02% were committed on go trials (omission errors). Similarly, response times were longer for correct than incorrect trials (M = .35 seconds and SD = .06 seconds for correct, M = .21 seconds and SD = .05 seconds for incorrect, and a t(68) = -17.36, p < .001).

Table 1

Intercorrelations of behavioural and self-report data

|  | 1 | 2 | 3 | 4 |
|---|---|---|---|---|
| SES | - | | | |
| CESD | -.46** | - | | |
| Incorrect response rt | .19 | -.10 | - | |
| GNG Error Rate (%) | -.37** | .27 | .02 | - |
| Correct response rt | .34** | -.19 | .34* | -.25 |

GNG= Go / No-Go task    *p < .05, **p < .01

## ΔERN

A visual inspection of Figure 1 shows a negative deflection peaking approximately between -80 and 100 ms for the ERN, CRN and ΔERN. ERN conforms to a typical fronto-central topography. Overall, these spatio-temporal properties of our ERN data are consistent with a vast body of previous research (Amodio et al., 2007; Olvet & Hajcak, 2008; Pfabigan et al., 2013; Ruchensky et al., 2020). As expected, ERN mean amplitudes (M = -.17, SD = 2.96 were more negative than CRN mean amplitudes (M= 0.84, SD = 1.98), t(68) = -3.05, p =.003.

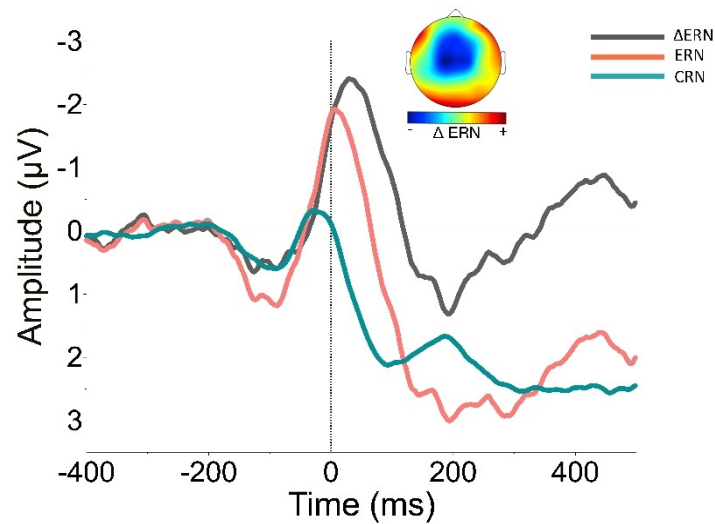

**Figure 1**. ERP waveforms for correct (CRN) and error (ERN) trials, and the difference wave (ΔERN) plotted on electrode Fz. Amplitude in microvolts (μV) is on the y axis and time in milliseconds is on the x axis. Two-dimensional scalp map plotting ΔERN mean amplitudes (-80 to 100 ms, min = -2.5μV to max = 2μV).

In order to examine whether the ΔERN was related to SES and CESD, we computed a number of correlations involving ΔERN (Table 2 and Figures 2, also see Table S1). Consistent with previous literature about the relationship between the ERN and psychopathology (Hajcak & Simons, 2002), we found a significant correlation between CESD and ΔERN (r = -.27, p < .001, see Figure 2d). Importantly, we also found a significant correlation between SES and ΔERN ($r$ = .40, $p$ <.001), which indicates that a larger ERN is observed in individuals with a lower SES (see Figure 2a). Although we analysed SES as a continuous variable in our main analysis, we also verified that our *a priori* SES groups also had significant differences in ERN amplitude (High-SES: *M*=-.18, *SD*=2.4; Low-SES: *M*=-1.8, *SD*=2.9, *t* = -2.58, *p* = .012). This finding is represented in figure 2c. Further details about the relationships between the study's main variables is available in the supplementary section (Supplementary Figures S1 and S2).

Table 2

Intercorrelations of study variables

|  | 1 | 2 | 3 |
|---|---|---|---|
| SES (Family Income) | - | | |
| ΔERN | .40** | - | |
| CESD | -.46** | -.27* | - |
| ΔPe | -.30* | -.22 | .11 |

*p < .05, **p < .01

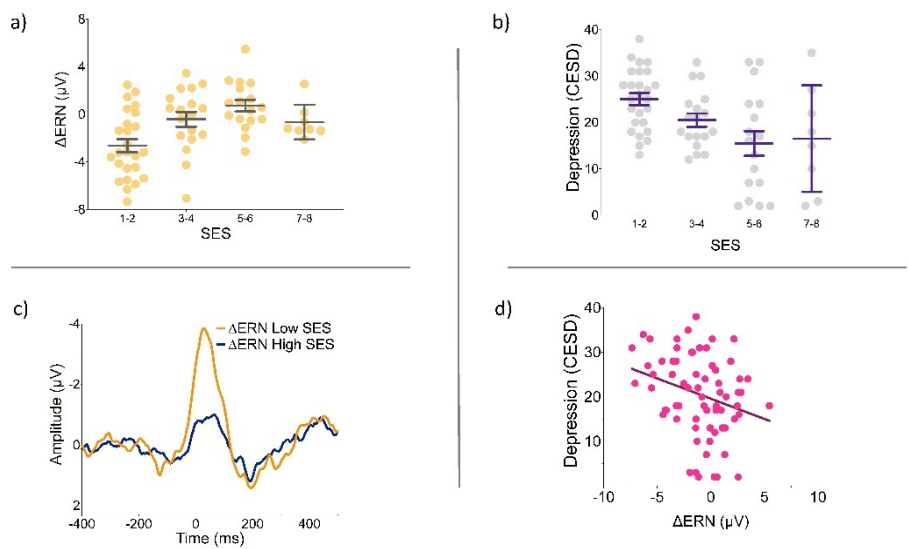

**Figure 2a.** Relationship between SES and ΔERN. The *x* axis represents increasing income categories described in the Participants section. Error bars represent +/-1 Standard errors of the mean. **2b**. Relationship between SES and CESD scores. **2c**. ERP waveforms plotted on electrode Fz for ΔERN separated according to a priori SES groups (low vs high SES, see "Participants" section). **2d**. Scatterplot describing the relationship between ΔERN and CESD scores.

In order to test whether SES scores accounted for the relationship between CESD and ΔERN, we ran a two-step hierarchical regression with ΔERN as the outcome variable.

In the first step, CESD was the sole predictor and in the second step we added SES. Given previous evidence on potential Gender effects in the relationship between psychopathology and the ERN (Ip et al., 2019), we included Gender in both steps.

As we show in Table 3, in the first step CESD was a significant predictor of ΔERN. However, adding SES in the second step led to a significant increase in the proportion of variance accounted for (ΔR$^2$= .10, $p$=.007) and CESD is no longer significant in this step, which indicates that the effect of CESD is accounted for by SES. This finding is confirmed by a mediation analysis based on 5,000 bootstrap samples showing that SES fully mediates the relationship between CESD and ΔERN (Bootstrap estimate = -.05, 95% CI [-.097, -.015]). This mediation was also significant for individual ERN (95% CI: -.10, -.01), but not for the CRN (95% CI: -.02, .03) waveforms. A moderation analysis showed that the interaction between SES and CESD did not reach significance levels for ΔERN, ERN or CRN (all ps>.15). Both Mediation and moderation analysis used the PROCESS macro (Hayes, 2012). We show in the supplementary section a breakdown of hierarchical regression results for the individual ERN and CRN waveforms (See Tables S2, S3).

Table 3

Hierarchical regression for ΔERN

| | B | SE | $\beta$ | $t$ | R | R$^2$ | ΔR$^2$ |
|---|---|---|---|---|---|---|---|
| **_Step 1_** | | | | | .37 | .11 | .11* |
| CESD | -.08 | .04 | -.27 | -2.35* | | | |
| Gender | -1.05 | .68 | -.18 | -1.54 | | | |
| **_Step 2_** | | | | | .45 | .20 | .10* |
| CESD | -.03 | .04 | -.11 | -.88 | | | |
| Gender | -1.01 | .65 | -.17 | -1.56 | | | |
| SES | .43 | .15 | .35 | 2.79** | | | |

Dependent variable = ΔERN. *p < .05, **p < .01, ***p < .001

These results indicate that SES is a strong predictor of ERN amplitude and that it fully accounts for the relationship between CESD and ΔERN. However, it could be argued that examining CESD and SES as continuous variables may not capture whether the actual risk of becoming severely depressed and the risk of living in poverty might both have unique effects on ΔERN. In order to address this possibility, we decided to analyse our data using a cumulative risk (CR) approach. CR models are thought to have several advantages over classical OLS models and they typically operationalize risk as an extreme level of the distribution of a given variable (Evans et al., 2013). Following standard practice in CR research (e.g. Gerard & Buehler, 2004) both SES and CESD were dichotomized in that the 75th percentile of CESD scores and the 25th percentile of SES were assigned a score of 1. This approach yielded two binary variables reflecting a high risk of either depression or poverty. In addition, we were also able to obtain a cumulative index of these two risks by summing up these two variables.

We were then able to assess whether the concomitant presence of these two risks had a unique predictive effect on ΔERN. We then computed a hierarchical regression analysis in which Gender and risk for depression were entered in the first step, risk of poverty was entered in the second step and a binary cumulative risk index (in which the concomitant presence of both risks is assigned a value of 1) was entered in the final step.

As shown in table 4, the risks of poverty remain the most powerful predictor of ΔERN. Although the risk of depression was significant in step 1 ($\beta$= -.25, p=.03), it was no longer significant after risk for poverty was added to the equation in step 2. The concomitant presence of both risks does not significantly add explained variance to the model in step 3 ($\Delta R^2$= .0003, p=.87). These results indicate that SES accounts for the CESD-ΔERN relationship even when CESD and SES are operationalized using a method that emphasizes the extreme ends of their distributions.

**ΔPe**

Although we did not have specific hypotheses about the Pe, we explored the relationships between ΔPe and both SES and CESD for the sake of completeness as this potential is often associated with the ERN (Falkenstein et al., 2000). Consistent with previous literature, we did find that ΔPe was more positive for errors (M = .83, SD = 4.34) than for correct responses (M = -1.49, SD = 2.33, t(68) = 2.09, p = .04). ERP waveforms relative to the Pe are presented in Figure 3. However, ΔPe did not correlate with CESD (r = .11, p = .37). We did find a significant correlation between ΔPe and SES (r = -.30, p = .01) showing that low-SES individuals tend to have higher ΔPe amplitude than high-SES (See Figure 3). This correlation was driven by the Pe to correct, rather than incorrect responses (See table S1). Additional information about this data can be found in the supplementary materials.

Table 4

Cumulative Risk Hierarchical regression

|  | B | SE | $\beta$ | $t$ | R | $R^2$ | $\Delta R^2$ |
|---|---|---|---|---|---|---|---|
| **_Step 1_** |  |  |  |  | .31 | .10 | .10* |
| Gender | -1.1 | .68 | -.18 | -1.6 |  |  |  |
| Depression Risk | -1.5 | .73 | -.25 | -2.2* |  |  |  |
| **_Step 2_** |  |  |  |  | .52 | .27 | .17*** |
| Gender | -1.0 | .62 | -.17 | -1.6 |  |  |  |
| Depression Risk | -.82 | .69 | -.13 | -1.2 |  |  |  |
| SES Risk | -2.4 | .62 | -.43 | -3.9*** |  |  |  |
| **_Step 3_** |  |  |  |  | .52 | .27 | .0003 |
| Gender | -1.0 | .63 | -.18 | -1.6 |  |  |  |
| Depression Risk | -.93 | 1.0 | -.15 | -.93 |  |  |  |
| SES Risk | -2.5 | .74 | .44 | -3.3*** |  |  |  |
| Cumulative Index | .22 | 1.3 | .03 | .16 |  |  |  |

Dependent variable = $\Delta$ERN. *p < .05, **p < .01, ***p < .001

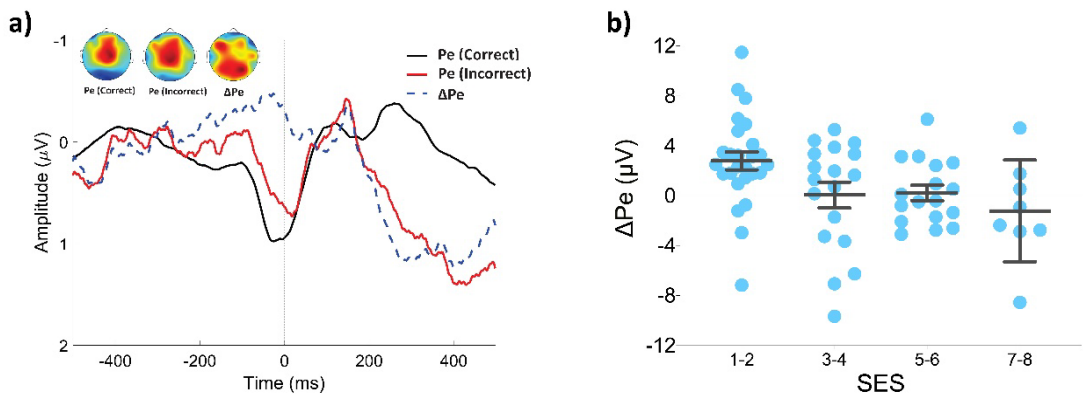

**Figure 3a.** ERP waveforms for the error positivity (Pe), plotted on an average of p3, Pz and P4 electrodes.  Amplitude in microvolts (µV) is on the y axis and time in milliseconds is on the x axis. Two-dimensional scalp map plotting Pe mean amplitudes (200-400 ms, min = -2.5µV to max = 2µV).**3b.** Relationship between SES and ΔPe. The x axis represents increasing income categories described in the Participants section. Error bars represent +/-1 Standard errors of the mean.

# DISCUSSION

To our knowledge, this is the first study to show that disparities in adult socioeconomic status (SES) are linked to the ERN, a well-known neural index of error monitoring. More specifically, we observed that adults living in low-income households had a larger ERN amplitude than people reporting higher incomes. This result was obtained while behavioral performance on error monitoring was better for higher than lower-income individuals. We also observed that ERN amplitude initially correlated with scores on the CESD, a well-known depression inventory. However, we found that this effect disappeared after SES was controlled for. Finally, we found that the error positivity (Pe) was higher for low-SES compared to high-SES and that Pe was not predicted by CESD scores. Hereafter we discuss the potential interpretation of these findings.

An important finding of our study was that the relationship between CESD scores and the ERN was not significant after controlling for SES. A potential explanation for this result could be that previous reports of a relationship between Depression and the ERN may have been driven by a relationship between the ERN and psychopathological states associated to depression (e.g. anxiety) but not necessarily captured by the CESD, rather than a direct relationship between depression and the ERN. Although we cannot formally rule out this possibility, it is highly unlikely: Even though the CESD is primarily designed to measure depression, it is also widely known to strongly correlate with a range of ID states that typically predict the ERN such as anxiety and obsessive-compulsive disorders (Abramowitz et al., 2009; Kavish et al., 2020; Orme et al., 1986). Although further research will be needed to replicate our study across different cultural and social contexts, our findings suggest that previous reports of an association between the ERN and psychopathological states may have partly been driven by factors related to socioeconomic differences. Our findings also tentatively suggest that existing contradictions in previous research about the relationship between the ERN and psychopathological states (Clayson et al., 2020) may be partly explained by a lack of consistent control of socioeconomic variables. Further, our results also suggest that caution is needed when considering the ERN as an index of psychopathological states, especially in low-income communities.

The precise nature of the mechanisms through which SES predicts ERN amplitude above and beyond psychopathology cannot be fully ascertained by our study because SES is a complex multi-dimensional construct. However, previous research suggests that individuals who live in a low-SES context may have biases in error monitoring because they would evaluate errors and losses as more consequential than higher SES individuals. This is supported by data showing that low-SES individuals face higher background risks in which errors and losses are inherently more consequential (Guiso & Paiella, 2008; Haushofer & Fehr, 2014), evidence that low-SES people tend to overreact to a broad range of aversive stimuli (see the review of Hao & Farah, 2020) , and fMRI evidence showing a higher reactivity in several brain areas to motivationally relevant stimuli for low-SES individuals (Gonzalez et al., 2016). It has been proposed that this tendency to overreact to motivationally-relevant stimuli may be related to a tendency of low-SES individuals to live in more punitive and uncertain environments (Gonzalez et al., 2016).  Therefore, it could be tentatively suggested that individuals living in a low-SES context may have error monitoring biases which could explain why they have a larger ERN amplitude than high-SES individuals. Further research is needed to fully establish the existence of such biases and how they may relate to psychopathological states.

Our results appear to be inconsistent with previous research indicating that low-SES children have a reduced ERN (Brooker, 2018; Conejero et al., 2016), and more generally, with models emphasizing that the ERN is an index of adaptive cognitive control mechanisms (Holroyd & Coles, 2002). The studies of Conejero et al. (2016) and Brooker (2018) are very different from our study regarding both overall methods and the age group of the participants and thus a direct comparison is not possible. However, evidence suggests that an exaggerated ERN reflecting error biases may be linked to learning mechanisms (Riesel et al., 2019). Therefore, it is possible that error biases may appear at stages of cognitive development that had not been reached by the very young children tested by Brooker (2018) and Conejero et al. (2016).

Finally, it is important to acknowledge that the results of this study are constrained by the geographical and demographic characteristics of our sample, which consisted of adults living in urban communities of Kuala Lumpur in Malaysia. More research will be needed to confirm if these findings can be found in different settings. However, the socio-geographical characteristics of our sample also contribute to addressing a well-known imbalance in psychological research, where a majority of studies rely on samples coming from Western, Educated, Industrialized, Rich, Democratic (WEIRD) countries (Henrich et al., 2010). This bias is particularly important for studies focusing on socioeconomic disparities (with some notable exceptions see Fernald et al., 2011; Mani et al., 2013), as developing countries tend to have more extreme social inequalities than WEIRD countries.

In summary, this study is the first to show a relationship between SES and ERN amplitude in adults. Specifically, we recorded larger ERN amplitudes from low-SES individuals living under the poverty line compared to higher SES individuals. We also found evidence that a correlation between the ERN and scores on a well-known depression inventory was fully accounted for by SES. These results indicate that neural responses to errors are exaggerated in low-SES individuals . These findings also highlight the importance of taking into account SES when using the ERN as a tool to measure psychopathological risks.

**ACKNOWLEDGMENTS**

This study was supported by the Universiti Kebangsaan Malaysia DCP-2017-014/1 grant.

**Supplementary Tables**

Table S1
Intercorrelations of all study variables

|  | 1 | 2 | 3 | 4 | 5 | 6 | 7 |
|---|---|---|---|---|---|---|---|
| SES | - | | | | | | |
| ERN | .39** | - | | | | | |
| CRN | .02 | .44** | - | | | | |
| Pe (incorrect) | -.13 | -.13 | -.01 | - | | | |
| Pe (correct) | .26* | -.01 | -.17 | .45** | - | | |
| ΔERN | .40** | .76** | -.25* | -.13 | .11 | - | |
| ΔPe | -.30* | -.14 | .09 | .84** | -.09 | -.22 | - |
| CESD | -.46** | -.32** | -.09 | .02 | -.14 | -.27* | .11 |

*$p < .05$, **$p < .01$

Table S2
Hierarchical Regression for ERN

|  | B | SE | β | t | R | $R^2$ | $\Delta R^2$ |
|---|---|---|---|---|---|---|---|
| **Step 1** | | | | | .37 | .14 | .14* |
| CESD | -.10 | .04 | -.32 | -2.8* | | | |
| Gender | -1.23 | .72 | -.19 | -1.71 | | | |
| **Step 2** | | | | | .45 | .21 | .07* |
| CESD | -.06 | .04 | -.18 | -1.45 | | | |
| Gender | -1.19 | .69 | -.18 | -1.72 | | | |
| SES | .39 | .16 | .29 | 2.4* | | | |

Dependent variable = ERN. *$p < .05$

Table S3
Hierarchical Regression for CRN

|  | B | SE | $\beta$ | t | R | $R^2$ | $\Delta R^2$ |
|---|---|---|---|---|---|---|---|
| **Step 1** |  |  |  |  | .10 | .01 | .01 |
| CESD | -.02 | .02 | -.09 | -.80 |  |  |  |
| Gender | -.18 | .51 | -.04 | -.35 |  |  |  |
| **Step 2** |  |  |  |  | .11 | .01 | .00 |
| CESD | -.02 | .03 | -.11 | -.83 |  |  |  |
| Gender | -.18 | .52 | -.04 | -.35 |  |  |  |
| SES | -.03 | .12 | -.03 | -.27 |  |  |  |

Dependent variable = CRN. *$p < .05$

**Supplementary Figures**

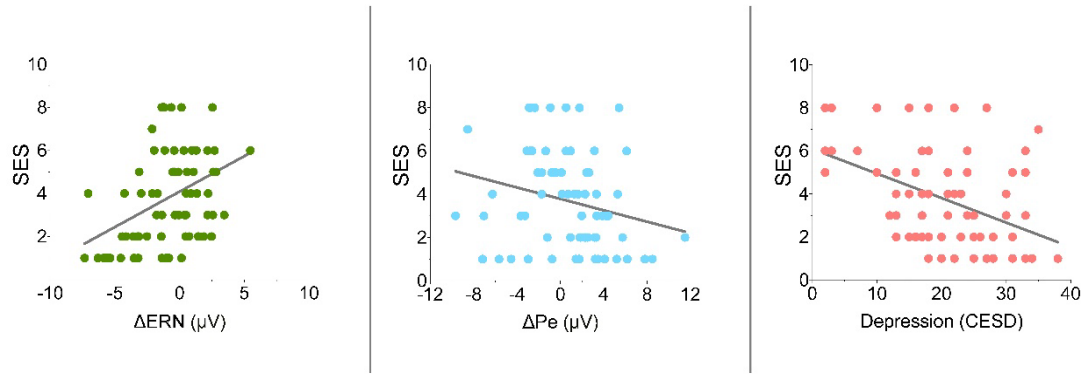

**Supplementary Figure S1.** Scatterplots describing the relationships between SES with  ΔERN, ΔPe and CESD.

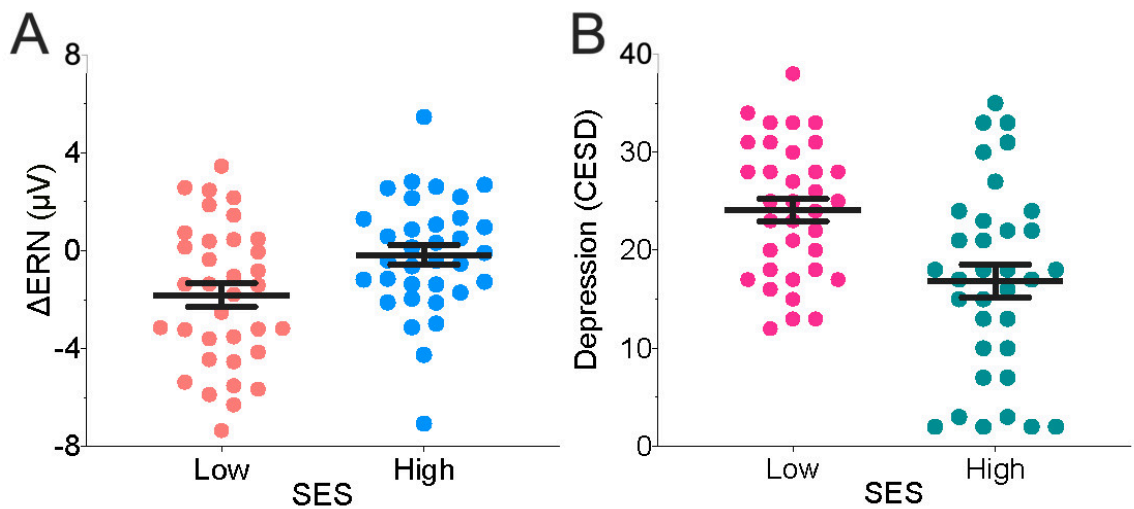

**Supplementary Figure S2.** Relationships between SES, ΔERN and CESD when SES is treated as a dichotomized variable (on the basis of *a priori* groups).

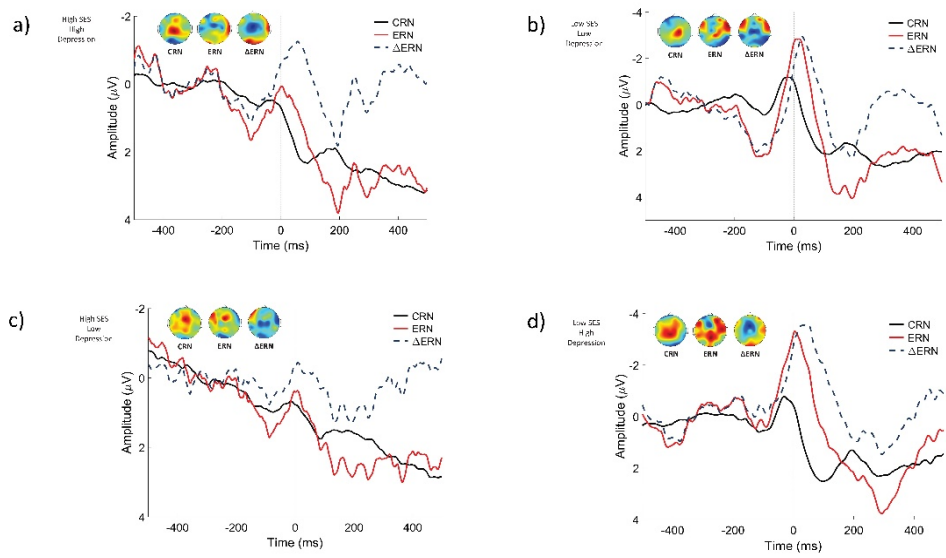

**Supplementary Figure S3.** ERN, CRN and ΔERN waveforms and scalp maps separately for subgroups of high-low SES and CESD.